\documentstyle[aps,psfig]{revtex}
\begin{document}
\date{\today}
\draft
\title{
Monte Carlo Study of the Anisotropic Heisenberg\\
Antiferromagnet on the Triangular Lattice}
\author{W. Stephan\thanks{Present address: Bishop's University, Lennoxville, 
Qu\'{e}bec, Canada J1M 1Z7} and B.W. Southern}
\address{
Department of Physics and Astronomy\\
 University of Manitoba \\
Winnipeg, Manitoba, Canada R3T 2N2}
\maketitle
\begin{abstract}
   We report a Monte Carlo study of the classical antiferromagnetic
   Heisenberg model with easy axis exchange anisotropy on the    
   triangular lattice. 
   Both the free energy
   cost for long wavelength spin waves as well as for the formation 
   of free vortices are obtained from the spin stiffness and
   vorticity modulus respectively. Evidence for two distinct 
   Kosterlitz--Thouless 
   types of 
   defect-mediated phase transitions at finite temperatures is presented.

\end{abstract}
\vspace{0.5in}
\pacs{PACS numbers: 75.10.Hk, 75.40.Mg}
\begin{twocolumn}
\section {INTRODUCTION}

    The nature of phase
transitions in frustrated systems are generally found to be quite
different from those in conventional magnets.\cite{1}   
Frustration often leads to non-trivial ground state degeneracies as in the case
of the antiferromagnetic spin 1/2 Ising model on the triangular lattice.
\cite{2,3}
The classical Heisenberg model on the two
dimensional triangular lattice with antiferromagnetic
nearest neighbour coupling and easy axis exchange anisotropy is another 
example where 
frustration leads to a novel ground state degeneracy.  Miyashita and 
Kawamura\cite{4}(MK)
have investigated both the ground state and the nature of harmonic 
excitations at low temperatures in this model.
 At zero temperature
the spins lie in a plane which contains the easy $z$-axis. In addition to 
the $S_1$
degeneracy related to the rotation of this plane about the $z$-axis, there 
is also a non-trivial $S_1$ degeneracy of the ground state related to 
rotations of
the spins within the plane. The planar spin configuration distorts as it is 
rotated
about an axis normal to the plane but the energy remains constant. This 
degeneracy is not
broken by the harmonic excitations. Using Monte Carlo methods, MK measured
the specific heat as well as various susceptibilities and
suggested that the system undergoes two successive phase transitions as
the temperature is lowered. The two transitions indicated the onset of
power law correlations of the spin-spin correlation functions parallel
and perpendicular to the easy axis. More recently, Sheng and Henley\cite{5} 
have examined
both the effects of finite temperature and quantum fluctuations on this 
non-trivial
degeneracy. They predicted that the continuous degeneracy at finite 
temperatures $T$
is reduced to a discrete
six-fold degeneracy and that an additional phase transition from a floating 
phase
to a locked phase occurs at a temperature in between the upper and lower 
transitions for large values of the easy axis anisotropy.

In the present work, we use a direct method to study the free energy cost of 
both spin wave and
vortex formation in this system at low temperatures. We define a spin
stiffness and a vorticity modulus in terms of an
equilibrium correlation function which can be evaluated using standard
Monte Carlo methods. Our results indicate that there are only two transitions
and that both correspond to vortex unbinding transitions accompanied by 
power law decay of spin correlations.

\section {The Model}

 The Hamiltonian of the system is given by
\begin{eqnarray}
 H = + \sum_{i<j} [{J^x S_i^xS_j^x + J^y S_i^yS_j^y+ J^z S_i^zS_j^z}] 
\end{eqnarray}
where ${ S^\alpha_i , \alpha =x,y,z}$ represents a classical 3-component 
spin of unit magnitude
located at each site $i$ of a triangular lattice and the exchange 
interactions $J^x=J^y=J, J^z=AJ$ are
restricted to nearest neighbour pairs of sites. We shall consider the case 
where the parameter range A $\ge$ 1 represents
an easy axis anisotropy. The limit $A\rightarrow 1$ corresponds to the 
isotropic Heisenberg model\cite{6} whereas the limit $A \rightarrow \infty$ 
corresponds to an infinite
spin Ising model.\cite{7}
 Southern and Xu\cite{8} have previously studied the Heisenberg limit and
found direct evidence for a vortex unbinding transition at finite temperature 
even
though the spin stiffness vanishes on large length scales. This latter fact 
indicates that the spin correlations
decay exponentially at all finite temperatures $T$. This behaviour
is quite different from that which occurs in the corresponding two-component 
$xy$ model\cite{9} where power law
decay of the correlations appear at low temperatures.

The triangular lattice can be decomposed into three sublattices $A,B$ and $C$
 as shown in figure 1
with the three spins on each triangle labelled as  $ {\bf S}_A, {\bf S}_B$
 and $ {\bf S}_C$. 
A chirality vector\cite{10} for each upward pointing triangle is defined as 
follows
\begin{eqnarray}
{\bf K_{\Delta}} = \frac{2}{3\sqrt{3}}({\bf S}_A \times {\bf S}_B + 
{\bf S}_B \times {\bf S}_C + {\bf S}_C \times {\bf S}_A)
\end{eqnarray}
\begin{figure} 
\centerline{\psfig{file=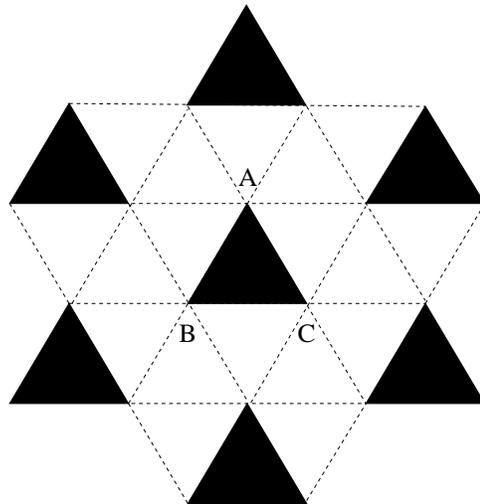,width=2.5truein}}
\caption{The decomposition of the triangular lattice into three sublattices 
$A,B$ and $C$. The sublattice labels on all black triangles are the same and 
represent only one of the six possible choices.}
\end{figure}
In the present case of the easy axis
antiferromagnet, the low temperature spin configuration
on each triangle has the spins lying in a plane which includes
the $z$-axis and which is perpendicular to the chirality vector.
We introduce
a local spin space coordinate system for each upward pointing triangle using 
the three unit vectors
$\hat{\bf K}_{\Delta}, \hat{\bf z}$ and  
$\hat{\bf{ \perp}}_{\Delta}=\hat{\bf z} \times \hat{\bf K}_{\Delta}$.
Hence the vectors $\hat{\bf z}$ and  $\hat{\bf \perp}_{\Delta}$ lie in the 
spin plane and $\hat{\bf K}_{\Delta}$ is the local normal to the plane. 
The $\hat{\bf z}$ is a global axis for all triangles whereas the 
$\hat{\bf K}_{\Delta}$ are only all aligned at $T=0$.
As $T$ increases, the local normal directions fluctuate and the spin plane 
develops curvature. Hence local coordinates are needed to properly define 
the response properties of the system.

Information about the rigidity of the system against fluctuations
 can be obtained from the spin wave stiffness coefficient.
The spin stiffness (helicity) tensor is given by the
second derivative of the free energy\cite{11,12} with respect
 to the twist angle about
a particular direction in spin space.  We apply a twist about each of the 
local axes defined
above and the corresponding average stiffness
is given by
\begin{eqnarray}
\rho_\alpha&=& -\frac{1}{N} \sum_{i<j} (\hat{\bf e}_{ij} \cdot
\hat{\bf u})^2 \langle J^\beta S_{i}^\beta S_{j}^\beta + 
J^\gamma S_{i}^\gamma S_{j}^\gamma
\rangle \nonumber\\
 & &-\frac{1}{NT} \langle ( \sum_{i<j} (\hat{\bf e}_{ij} \cdot
\hat{\bf u}) [\frac{J^\beta + J^\gamma}{2}][ S_i^\beta S_j^\gamma - 
S_i^\gamma S_j^\beta ])^2 \rangle 
\end{eqnarray}
where the superscipts take the values $\alpha=\perp_{\Delta},z,K_{\Delta}$ 
and $\alpha,\beta,\gamma$ are to be taken in cyclic
order. $S_i^\alpha$
denotes the component of the spin at site $i$ in the average direction of 
the corresponding
local unit vectors of the upward pointing triangles to which the spin 
belongs, $\hat{\bf e}_{ij}$ are unit vectors along neighbouring bonds
and $\hat{\bf u}$ is the {\it direction} of the twist {\it in the lattice}. 
  
 We employ compact clusters, as shown in figure 2, containing $N$ spins
with periodic boundary conditions applied that are consistent with the 
sublattice structure. The number of spins is related to the linear size $L$
of the cluster as $N=3L^2$. 
\begin{figure}
\centerline{\psfig{file=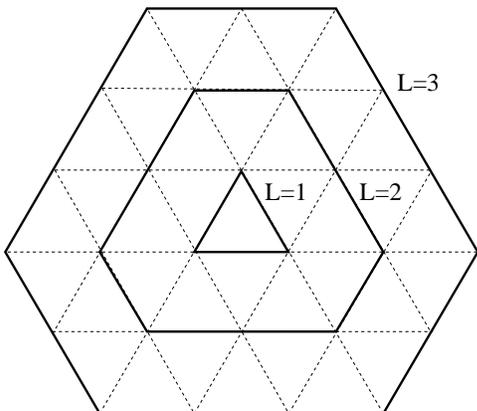,width=2.5truein}}
\caption{Compact clusters of linear size $L$ containing $N=3L^2$ spins used 
in the Monte Carlo calculations are indicated by the solid lines. In each 
case periodic boundary conditions are used which are compatible with the 
three sublattice structure.}
\end{figure}
We have used a single spin-flip heat bath algorithm\cite{13} to update the spin
directions at each Monte Carlo step and all thermal averages are replaced by
 time averages. For the largest value of the system size $L=60$ studied we
 discard
the first $1.5 \times 10^4$ Monte Carlo steps and perform averages over 
the next $5 \times 10^5$ steps. 

The values of the stiffnesses
at zero temperature can be evaluated using the exact classical ground 
state correlation functions. We find these values to be 
\begin{eqnarray}
\rho_z/{J} &=& \frac{2A +1}{2(1+A)^2}\nonumber \\
\rho_{\perp}/{J} &=& A^2 \frac{A +2}{2(1+A)^2}\nonumber \\
\rho_K &=&  \rho_z + \rho_{\perp}  
\end{eqnarray}
These three stiffnesses satisfy a perpendicular axis theorem which is 
consistent with the
ground state being a planar spin arrangement. In the Heisenberg limit we 
have $\rho_z = \rho_{\perp}$ but for large values of $A$ they have 
different energy scales, ${\rho_z}/{J}
\sim \frac{1}{A}$ whereas ${\rho_{\perp}}/{J} \sim \frac{A}{2}$. 
These limiting values can also be
obtained from the harmonic excitation spectrum.\cite{5}
\vspace{5mm}
\begin{figure}
\centerline{\psfig{file=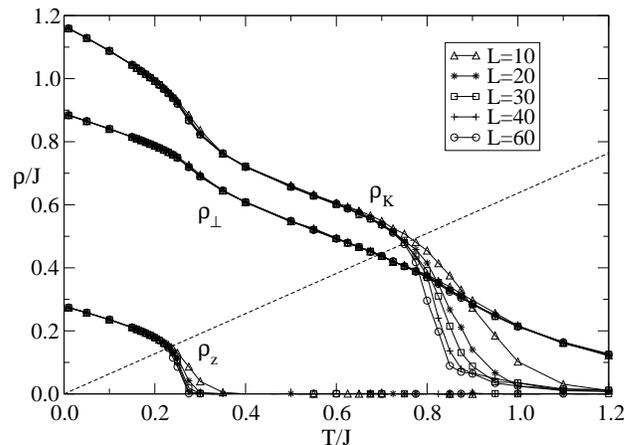,width=3.2truein,angle=270,clip=}}
\caption{The three stiffnesses $\rho_z, \rho_{\perp}$ and $\rho_K$ in units 
of $J$ as a function of $T/J$ for system sizes $L=10,20,30,40,60$ when $A=2$. 
The dashed curve represents the line $\rho = (2/\pi)T$.}
\end{figure}
Figure 3 shows the results obtained for the  three stiffnesses as a
function of $T/J$ for various system sizes $L$ with $A=2$. As the 
temperature increases
towards $T_{c_1} = (0.231 \pm 0.005) J$, the stiffness corresponding to the 
smaller energy scale, $\rho_z$,
decreases abruptly and  a strong dependence on system size $L$ is evident 
indicating that above this temperature the system has no
rigidity against twists of the spins about the easy axis.  MK suggested that
this transition corresponds to the unbinding of vortices associated with the
chirality vector which lies primarily in the $xy$ plane. We will provide 
further support for this interpretation in section IV.
 
The remaining two stiffnesses do not vanish at $T_{c_1}$ but do exhibit a 
rapid decrease 
followed by a more steady decrease until a higher temperature 
$T_{c_2} = (0.75 \pm 0.03) J$. 
At this higher temperature, the strong dependence of the stiffness $\rho_K$ 
on system size $L$ indicates that the system
loses rigidity against rotations about the local chirality axes. 
The Kosterlitz-Thouless(KT) theory\cite{14,15,16} for the $xy$ model in 
two dimensions predicts a universal value for the ratio ${\rho_c}/{T_c}$ 
equal to $2/ \pi$. The dashed line in figure 3 corresponds to  
$\rho /T = {2}/{ \pi}$ and intersects the stiffnesses at temperatures 
where finite size effects first appear.   
The stiffnesses $\rho_z$ and $\rho_K$
indicate two temperatures where rigidity of the system about the $z$-axis 
is first
lost followed by a loss of rigidity about the local chirality axis with an 
apparent universal value for the ratio $\rho_c /T_c$ in each case.  The 
third stiffness $\rho_{\perp}$ has a change in curvature at the two 
transitions but remains finite to high temperatures. Similar behaviour is 
found for
other values of $A$ as well. 
\begin{figure}
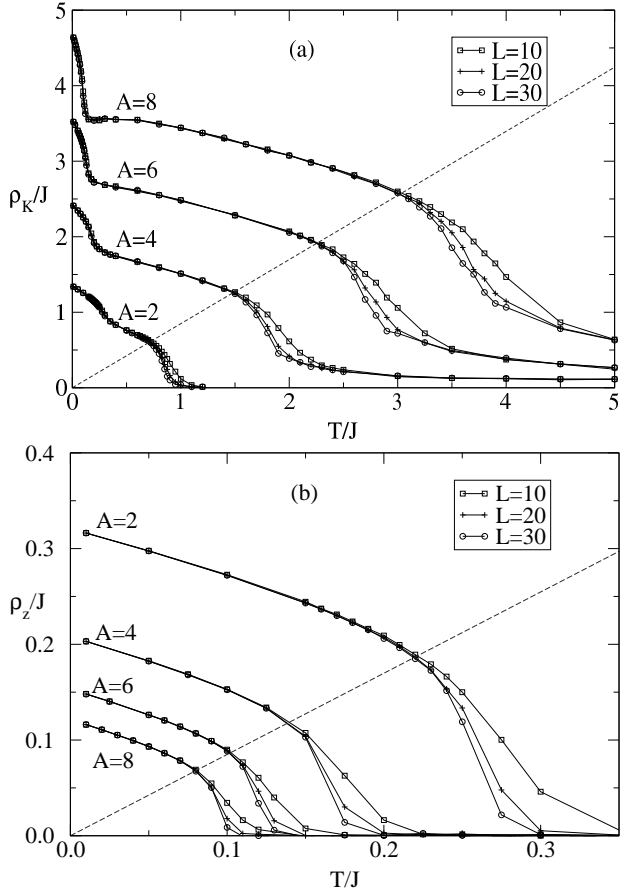

\centerline{\psfig{file=fig4a.eps,width=3.2truein,angle=270,clip=}}
\centerline{\psfig{file=fig4b.eps,width=3.2truein,angle=270,clip=}}
\caption{The spin stiffnesses (a) $\rho_K$  and (b) $\rho_z$ in units of $J$ 
as a function of $T/J$ for system sizes $L=10,20,30$
with $A=2,4,6,8$. The dashed curve in both cases represents the line 
$\rho = (2/\pi)T$.}
\end{figure}
Figures 4(a) and 4(b) show our results for 
$\rho_K$ and $\rho_z$ respectively for the values $A=2,4,6,8$. The dashed 
line plotted in both figures corresponds to $\rho /T = {2}/{ \pi}$ and in 
each case this line intersects the stiffnesses at the temperatures where 
finite size effects first become significant. These results indicate that 
the ratio ${\rho_c}/{T_c}$ appears to have
the universal value ${2}/{ \pi}$ for both stiffnesses for all values of 
$A>1$. We have used this criterion to estimate the values of $T_{c_1}$ 
and $T_{c_2}$.

In the same way that the spin stiffness is a measure of the response of
the spin system to a twist over the length of the lattice, a vorticity\cite{8}
can be defined as the response of the spin system to an imposed twist 
about a given axis $\alpha$ in spin space
along a closed path which encloses a vortex core. This is 
essentially the response of the system to an isolated vortex and can
be calculated as the second derivative of the free energy with respect
to the strength of the vortex, or winding number $m$, evaluated at $m=0$.
We obtain the following expression  
\begin{eqnarray}
V_\alpha&=& -\frac{\sqrt{3}}{4 \pi} \sum_{i<j} (\frac{\hat{\bf e}_{ij} \cdot
\hat{\bf \phi_i}}{r_i})^2 \langle J^\beta S_{i}^\beta S_{j}^\beta + 
J^\gamma S_{i}^\gamma
 S_{j}^\gamma
\rangle \nonumber\\
 & &-\frac{\sqrt{3}}{4 \pi T} \langle ( \sum_{i<j} (\frac{\hat{\bf e}_{ij} 
\cdot
\hat{\bf \phi_i}}{r_i}) [\frac{J^\beta + J^\gamma}{2}][ S_i^\beta S_j^\gamma 
- S_i^\gamma S_j^\beta ])^2
 \rangle  
\end{eqnarray}
where $r_i$ is the distance of site $i$ from the vortex core and $\hat{\bf
 \phi_i}$ is tangent to the circular {\it path in the lattice}
passing through the site $i$ and
enclosing the vortex. Here $\alpha, \beta, \gamma$ are defined in the same way
as for the stiffnesses and indicate the axis of rotation of the vortex.

The $V_{\alpha}$ contain both a core contribution and a part which is
proportional to ln$L$. By comparing systems of different lattice sizes 
$L_1$ and $L_2$ we
can extract the vorticity modulus $v_{\alpha}$ defined as follows
\begin{eqnarray}
 V_{\alpha} =  C_{\alpha} +  v_{\alpha} \ln L
\end{eqnarray}
using 
\begin{eqnarray}
 v_\alpha &= & \frac{V_\alpha (L_2) - V_\alpha (L_1)}{\ln (L_2/L_1)}
\end{eqnarray}
where the $v_\alpha$ are normalized so that they have the same zero 
temperature values as the corresponding stiffnesses.
Our
approach does not require any change in boundary conditions and is applied
directly to the antiferromagnetic model.

\begin{figure}
\centerline{\psfig{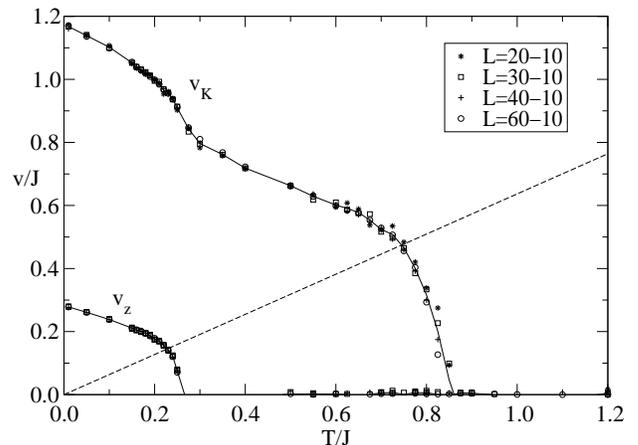}}
\caption{The vorticity moduli $v_K$ and $v_z$ in units of $J$ as a function 
of $T/J$ obtained using various pairs of system sizes when $A=2$. The solid 
curves represent the average of all pairs for each modulus. The dashed curve 
represents the line $v = (2/\pi)T$.}
\end{figure}
Figure 5 shows our results for the vorticity moduli obtained by comparing 
systems of different sizes when $A=2$. The behaviour of the moduli as a 
function of increasing temperature is identical to the spin stiffnesses 
shown in figure 3. The vorticity modulus $v_{\perp}$ is only weakly 
sensitive to the two transitions and is not plotted.
The vanishing of the vorticity moduli $v_z$ and $v_K$ at $T_{c_1}$ and 
$T_{c_2}$ respectively indicates that free vortices appear at these two 
transition temperatures. The results are also consistent with a universal 
value $v_c/T_c = 2/\pi$. Similar behaviour is found for the values 
$A= 2,4,6,8$. 

The results presented so far do not indicate directly that power law decay 
of correlations are present below these two transition temperatures. 
In the next section we present results for
the spin-spin structure factor which are also consistent with KT transitions.

\section {Structure Factor}

 We have also studied the spin-spin structure factor for various system 
sizes $L$
\begin{eqnarray}
S^{\alpha\alpha}({\bf q}) = \frac{1}{N} \sum_{i,j} <S^\alpha_i S^\alpha_j> 
e^{i {\bf q} .
( {\bf r}_i - {\bf r}_j )}
\end{eqnarray}
where $\alpha = x,y,z$ . In particular, we have studied 
$S^{\perp}=S^{xx}+S^{yy}$ and $S^{zz}$ as a function of both ${\bf q}$ 
and $T$. In both cases the Fourier component ${\bf Q} = (4\pi /3,0)$  
exhibits a divergence at and below the two transition temperatures $T_{c_1}$ 
and $T_{c_2}$ respectively. Other values of ${\bf q}$ only exhibit a maximum. 
This Fourier component corresponds to the three-sublattice structure 
associated with the triangles. If we assume power law decay of the spin-spin 
correlation functions, 
then the structure factors should depend on the system size $L$
as follows
\begin{eqnarray}
S^{zz}({\bf Q})&\sim& L^{2-\eta_z} \nonumber \\
S^{\perp}({\bf Q}) &\sim& L^{2-\eta_{\perp}}
\end{eqnarray}
where $\eta_z$ and $\eta_{\perp}$ are the corresponding correlation length 
exponents.

\begin{figure}
\centerline{\psfig{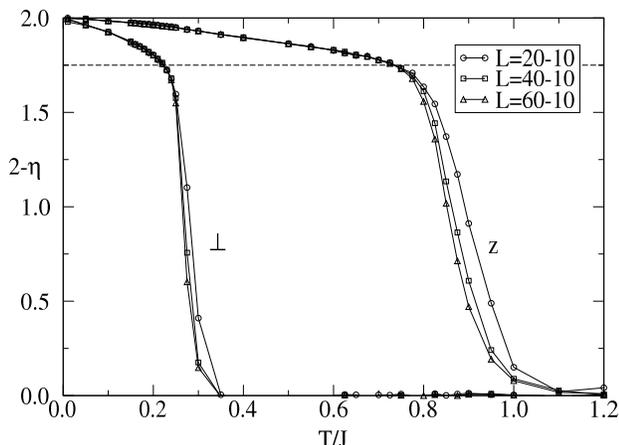}}
\caption{The correlation function exponents $2-\eta_{\perp}$ and $2-\eta_z$ 
as a function of $T/J$ obtained by comparing various system sizes with $A=2$. 
The dashed line corresponds to $\eta=1/4$.}
\end{figure}
Figure 6 shows the values of $2-\eta_z$ and $2-\eta_{\perp}$ as a function of 
$T/J$ obtained by comparing systems of different sizes with $A=2$.
The dashed line indicates the value $\eta=1/4$ predicted by the KT theory and 
intersects both exponents at temperatures where a strong dependence on system 
size first appears. Similar results were obtained for other values of $A$.
The universal value of $\eta_z=\eta_{\perp}=1/4$ is consistent with the
universal value of $\rho_c/T_c = 2/\pi$ and $v_c/T_c = 2/\pi$ obtained in 
section II. 

\begin{figure}
\centerline{\psfig{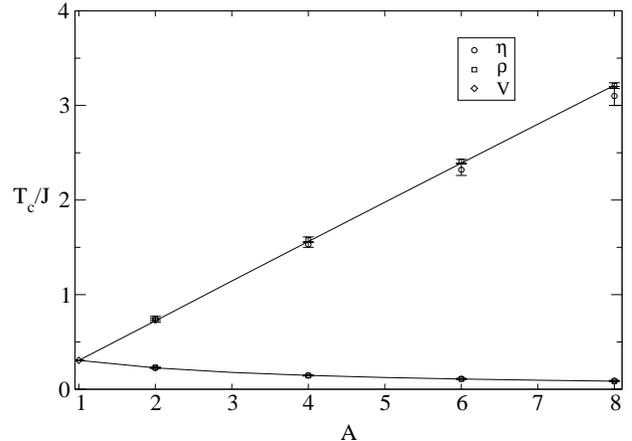}}
\caption{The transition temperatures $T_{c_1}/J$ (lower) and $T_{c_2}/J$ 
(upper) as a function of $A$ determined using the universal value of the 
stiffness $\rho$ and the universal value of $\eta$ are shown by the squares 
and circles respectively. The value at the Heisenberg point ($A=1$) is 
indicated by the diamond and was obtained from previous vorticity results. 
The solid
curves are obtained by assuming that these two temperatures are proportional 
to the $T=0$ values of $\rho_z$ and $\rho_{\perp}$ respectively.}
\end{figure}
We have used these two criteria to obtain the temperatures $T_c{_1}$ and 
$T_c{_2}$ as a function of $A$ as shown in figure 7. The values obtained 
from the stiffnesses and the structure factors agree with each other fairly 
well. 
The error bars in $T_c$ reflect contributions from the statistical 
uncertainty of the Monte Carlo measurements, as well as estimated systematic 
contributions due to finite size effects. 
If we make the simple assumption that the dependence of $T_{c_1}$ and 
$T_{c_2}$ on $A$ is the same as the zero temperature values of $\rho_z$ 
and $\rho_{\perp}$ respectively, then we have
\begin{eqnarray}
T_{c_1} &=& T_{c}\ \frac{4 (2A+1)}{3(1+A)^2} \nonumber \\
T_{c_2} &=& T_{c}\ \frac{4 A^2 (A+2)}{3(1+A)^2}
\end{eqnarray}
where the values of $T_{c_1}$ and $T_{c_2}$ in the isotropic limit $A=1$ are 
both equal to $T_{c}=(.305 \pm .005)J$.\cite{8} These expressions are plotted 
as the solid curves in figure 7 and the agreement with the data points is 
remarkable. 

The limit $A \rightarrow \infty$ of the present model is equivalent to the 
$S \rightarrow \infty$ limit of the antiferromagnetic spin $S$ Ising model 
on the same lattice with coupling $AJ$. The solid curve for $T_{c_2}$ 
approaches the value  $T_c{_2}/AJ \sim .407 \pm .007$ in this limit. 
This value is also in excellent agreement with that obtained by Nagai et 
al.\cite{17} for the $S=\infty$ Ising model. 
In the next section we give a more detailed microscopic description of 
these phases.

\section {Microscopic Properties}
The three spins $ {\bf S}_A, {\bf S}_B$ and ${\bf S}_C$ on each triangle
can be expressed in terms of the following three vectors
\begin{eqnarray}
{\bf M}_0 &=& {\bf S}_A + {\bf S}_B + {\bf S}_C \nonumber\\
{\bf M}_R &=& {\bf S}_A -\frac{1}{2} {\bf S}_B -\frac{1}{2} {\bf S}_C 
\nonumber\\
{\bf M}_I &=& \frac{\sqrt{3}}{2}({\bf S}_B - {\bf S}_C)
\end{eqnarray}
The vectors ${\bf M}_R$ and ${\bf M}_I$ are the real and imaginary parts 
of the complex Fourier
component ${\bf M}_{\bf Q}= {\bf M}_R + i {\bf M}_I$ where 
${\bf Q} = (4\pi /3,0)$. In addition, the chirality for
each triangle can be written as 
${\bf K} = \frac{4}{9} {\bf M}_R \times {\bf M}_I$.

In the ground state the three spins lie in a plane which includes the 
$z$-axis. We denote
the angle  that each sublattice spin makes
with the $z$-axis by $\theta_\alpha $, and define
\begin{eqnarray}
\psi = (\theta_A + \theta_B + \theta_C)/3
\end{eqnarray}
where $0 \le \psi \le 2\pi$. We measure $\psi$ relative to the $z$-axis 
in the counter-clockwise direction. As discussed by Sheng and Henley\cite{5}, 
the nontrivial degeneracy of the ground state can be parameterized
in terms of the angle $\psi$. It has been previously
pointed out that both the energy and the magnitude of ${\bf M}_0$ are 
constant in the ground state manifold\cite{5,18}.
However, it is easy to show that the modulus of the complex vector 
${\bf M}_{\bf Q}$ is also
constant. In the Appendix we give some exact results for the components 
of these vectors in the ground state. The $\perp,z$ components of these 
two vectors depend on $\psi$ in quite 
different ways. The components of ${\bf M}_0$ have periodicity $2\pi /3$ 
whereas the
vectors ${\bf M}_R$ and ${\bf M}_I$ have period $2\pi$. 

\begin{figure}
\centerline{\psfig{file=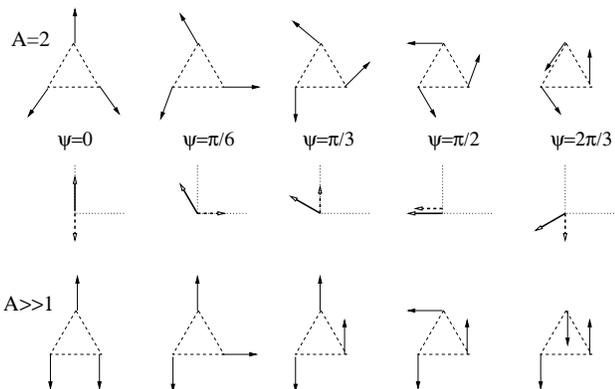,width=3.2truein,clip=}}
\caption{Ground state spin configurations for values of $\psi$ in the 
range $0 \le \psi \le 2\pi/3$ for $A=2$ and $A \gg 1$. The horizontal and 
vertical axes are $\perp$ and $z$ repsectively. The corresponding directions 
of ${\psi}$ and ${\bf M}_0$ are independent of the value of $A$ and are 
indicated in the middle row by the solid and dashed arrows respectively.}
\end{figure}
Figure 8 shows ground state spin configurations of the sublattices in the 
$\perp,z$ plane for values of $\psi$ in the range $0 \le \psi \le 2\pi/3$. 
The top row corresponds to $A=2$ and the bottom row to $A \rightarrow \infty$. 
Also shown are the corresponding directions of ${\psi}$ and ${{\bf M}}_0$ 
which are independent of the value of $A$. Rotations of the angle $\psi$ by 
$2\pi/3$ correspond to interchanges of the sublattice spin directions and to 
a complete rotation of ${\bf M}_0$ by $2\pi$. Hence vortex-like 
configurations of ${\bf M}_0$  simply correspond to switches of the 
sublattice magnetizations. For large values of the easy axis anisotropy, 
the spin configurations become more Ising-like. In the range 
$0 \le \psi \le \pi/3$, sublattices $A$ and $B$ remain locked parallel to 
the easy axis while sublattice $C$ simply follows  ${\bf M}_0$. In the 
range $\pi/3 \le \psi \le 2\pi/3$, the roles are interchanged with 
sublattice $A$ following the direction of ${\bf M}_0$. These sublattice 
changes as a function of $\psi$ are similar to the domain transition 
regions described by Wannier\cite{2} for the $S=1/2$ Ising model at $T=0$. 
In this latter case, the transition regions contribute to the macroscopic 
entropy of the system and to the power law decay of correlations with 
wavevector ${\bf Q}$. \cite{3}

The moduli of the complex numbers $M^z_{\bf Q}=M^z_R+ i M^z_I$ and 
$M^{\perp}_{\bf Q}=M^{\perp}_R + i M^{\perp}_I$ are not constant in the 
ground state but the combination
\begin{eqnarray}
M^z_{\bf Q}   +i M^{\perp}_{\bf Q}=\sqrt{\frac{(4A+2)(A+2)}{(1+A)^2}}
 e^{i\Psi}
\end{eqnarray}
has constant modulus and a phase angle $\Psi$
which has a leading term linear in $\psi$ but with an additional
small 6-fold modulation. At low temperatures the chirality vector lies in 
the $xy$ plane and it can be described by the complex number
\begin{eqnarray}
K_x + i K_y = \sqrt{K^2_x+K^2_y} e^{i\Theta}
\end{eqnarray}
where the modulus is not constant but exhibits only a weak $\cos(6\psi)$ 
variation in the ground state. These two phase angles $\Theta$ and $\Psi$ 
describe the complex order parameters associated with the transitions at 
$T_{c_1}$ and $T_{c_2}$ respectively.

Our Monte Carlo procedure permits us to take snapshots of the spin 
configurations at various temperatures. In order to separate the topological 
defects from the continuous deformations of the spin configurations, we 
first raise the temperature to some fixed value and allow the system to 
reach equilibrium. We then rapidly quench the system to low temperatures 
and allow the system to approach a nonequilibrium configuration. The 
topological defects are metastable at low temperatures and require a much 
longer time to disappear than the continuous deformations. 
In figures 9 and 10 we show snapshots of the spin configuration at very 
low temperatures obtained by heating the system above $T_{c_1}$ and 
$T_{c_2}$ respectively and then rapidly quenching. In both cases, (a) 
describes the spatial variation of the chirality angle $\Theta$ and (b) 
describes the angle $\psi$.

\begin{figure}
\centerline{\psfig{file=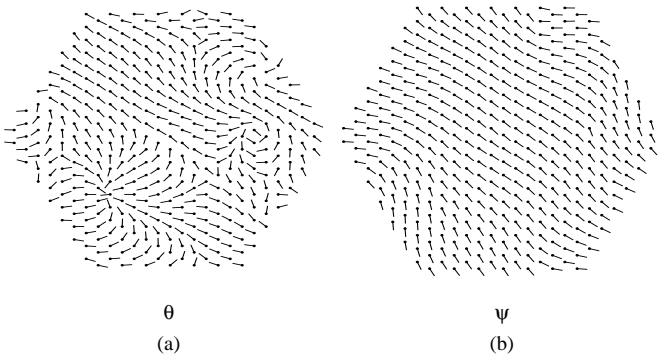,width=3.5truein,clip=}}
\caption{Snapshots of (a) the chirality angle $\Theta$ and (b) the angle 
$\psi$ at very low temperature following a quench from just above $T_{c_1}$ 
for $A=2$ and $L=20$. The dots indicate the positions of the black triangles 
shown in figure 1 and the lines indicate the local direction of either 
$\Theta$ or $\psi$.}
\end{figure}
Figure 9 shows snapshots of the spin configuration for a cluster of size 
$L=20$ with $A=2$ after quenching from a
temperature of $T=0.5 J$, just above $T_{c_1}$.  Vortices associated with
the angle $\Theta$ are clearly visible in (a) whereas only continuous 
distortions of the angle $\psi$ are visible in (b). These vortices in 
$\Theta$  only persist at low $T$ if the system
is first heated to a $T>T_{c_1}$ and then quenched. They are metastable 
configurations at this lower temperature but become stable at $T_{c_1}$. 
Our results provide direct evidence for a vortex unbinding transition at 
$T_{c_1}$ involving the chirality vector. 
\begin{figure}
\centerline{\psfig{file=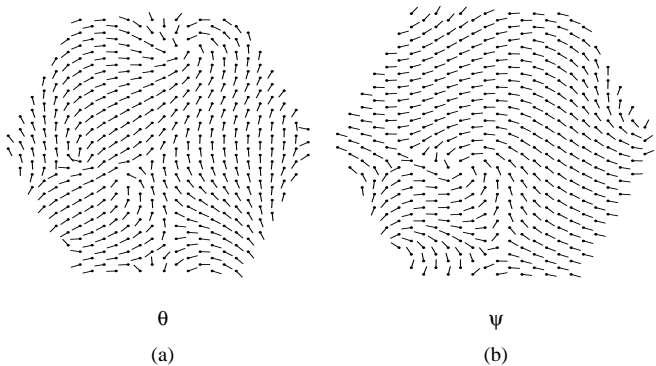,width=3.5truein,clip=}}
\caption{Snapshots of (a) the chirality angle $\Theta$ and (b) the angle 
$\psi$ at very low temperature following a quench from above $T_{c_2}$ for 
$A=2$ and $L=20$. The dots indicate the positions of the black triangles 
shown in figure 1 and the lines indicate the local direction of either 
$\Theta$ or $\psi$.}
\end{figure}
Figure 10 shows similar
snapshots when the system is first heated to a temperature $T=3J$, well 
above $T_{c_2}$, and then rapidly quenched. Vortices in both $\Theta$
and $\psi$ are clearly visible in this case. The vortices in $\psi$ are 
only visible after quenching from above $T_{c_2}$. Hence the upper 
transition corresponds to a vortex unbinding transition associated with the 
sublattice ordering vector ${\bf M}_{\bf Q}$.

We have also studied how the sublattice magnetizations change as we update 
the spin configurations. In the temperature range $ T < T_{c_2}$ there is a 
continuous sublattice switching that occurs.
The time scale for the switching depends both on $T$ and system size $L$ 
increasing with $L$ and $1/T$. We have not been able to determine
a scaling form for this time scale. Similar behaviour has been reported 
previously for other frustrated systems on the triangular lattice. The 
antiferromagnetic spin $1/2$ Ising model with both nearest and next-nearest 
neighbour interactions\cite{19,20} as well as the antiferromagnetic 
$S=\infty$ nearest neighbour model\cite{7,21,22} exhibit this phenomena in 
the temperature range where a KT phase occurs. In our case, the switching is 
due to the wave-like variations in $\psi$ which can
produce ``vortex-like'' variations in the local magnetization vector
${\bf M}_0$. In the range of $A$ that we have studied, we have not observed 
a transition from sublattice switching to a partially ordered phase where 
one of the sublattices is locked in the $+z$ direction, another in the 
$-z$ direction and the third perpendicular to the $z$-axis. Sheng and 
Henley\cite{5} predicted that such a locking transition might occur at 
low enough temperature.

\section {Summary}

We have presented strong evidence for the occurence of two distinct 
Kosterlitz-Thouless types of defect-mediated phase transitions in the 
Heisenberg antiferromagnet on the triangular lattice
with easy axis anisotropy.
A numerical approach was used to calculate the
rigidity of the system against both spin wave deformations and the formation 
of free vortices at low
temperatures for several values of the easy axis anisotropy. In each case 
a universal value of the stiffness or vorticity modulus, 
$\rho_c/T_c = v_c /T_c = 2/\pi$, was found. In addition, the spin 
correlation length exponent has the universal value $\eta=1/4$.
These are the same values that are predicted by the Kosterlitz-Thouless 
theory for defect unbinding
transitions in two dimensions.  The two phase angles identified in equations 
(13) and (14) in section IV describe the complex order parameters 
associated with the transitions at $T_{c_2}$ and $T_{c_1}$ respectively.

We find that the spin stiffnesses and the corresponding vorticity moduli 
behave identically for the easy axis case $A>1$. In contrast, at the 
isotropic Heisenberg limit, the spin stiffness vanishes at large length 
scales whereas the vorticity moduli are non-zero at low T and vanish 
abruptly at a finite temperature.\cite{8} Our previous work on the $xy$ 
model \cite{9} also indicated that the vorticity
and stiffness behave identically. This corresponds to the 
$A \rightarrow -\infty$ limit where there are again two transitions but
they are very close in temperature with the upper transition corresponding 
to an Ising-like transition and the lower to a Kosterlitz-Thouless 
transition. Recent work by Capriotti et. al. \cite{23} has also found 
similar behaviour in the range $0< A <1$. Hence the Heisenberg point is a 
multicritical point where four phase transition lines meet. For $A>1$ 
there are two KT transition lines
whereas as for $A<1$ there is an Ising and KT line.

\section{Acknowledgements}                                

This work was supported by the Natural Sciences and Engineering
Research Council of Canada.

\appendix
\section*{}

 Miyashita and Kawamura\cite{4} were the first to identify the nontrivial 
degeneracy of the ground state configuration. Other groups\cite{5,18} have 
studied the effects of quantum fluctuations
on this degeneracy. One indicator of the degeneracy is that the 
magnetization vector ${\bf M}_0$ has a constant magnitude in the ground 
state independent of the value of $\psi$. 
In particular we find the following
exact relations
\begin{eqnarray}
 M^{\perp}_0 &=& -(\frac{A-1}{A+1}) \sin(3 \psi) \nonumber\\
M^{z}_0 &=& -(\frac{A-1}{A+1}) \cos(3 \psi) \nonumber\\
{\bf M}_0 . {\bf M}_0 &=& (\frac{A-1}{A+1})^2 \nonumber\\
\end{eqnarray}
where $\psi$ is measured counter-clockwise from the $z$-axis in the 
$\perp,z$ plane.
A rotation of ${\bf M}_0$ by $2\pi$ corresponds to rotating $\psi$ by 
$2\pi/3$ which is simply a cyclic permutation of the sublattices on the 
triangle. Hence vortices in ${\bf M}_0$ can be associated with sublattice 
switching.

In addition to ${\bf M}_0$, the complex vector ${\bf M_Q}$ also has constant 
modulus,
\begin{eqnarray}
{\bf M}_{\bf Q} .{\bf M}_{\bf -Q}&=&{\bf M}_R . {\bf M}_R + {\bf M}_I . 
{\bf M}_I \nonumber\\  &=& \frac{4A^2+10A+4}{(1+A)^2} 
\end{eqnarray}

The moduli of the components of ${\bf M}_{\bf Q}$ as well, as the chirality, 
are not constant in the ground state but depend on $\psi$
as follows
\begin{eqnarray}
|M_Q^z|^2 &=& \frac{3A^2+6A}{(1+A)^2} +(\frac{A-1}{A+1})^2 \cos^2(3\psi) 
\nonumber\\
|M_Q^\perp|^2&=& \frac{6A+3}{(1+A)^2} +(\frac{A-1}{A+1})^2 \sin^2(3\psi) 
\nonumber\\
|{\bf K}|^2 &=& \frac{4}{3\sqrt{3}}[ \frac{\sqrt{2A+A^2}}{(A+1)^2}(2+A)
\cos^2(3\psi)\nonumber \\
& &\ \ \ \ \ \ +
\frac{\sqrt{1+2A}}{(1+A)^2}(1+2A)\sin^2(3\psi)] 
\end{eqnarray}


%
\end{twocolumn}

\end{document}